\documentclass[final,12pt]{article}

\usepackage[pdftex,pdftitle={A new theoretical superstoichiometric hard $TiN_x$ crystal},pdfauthor={Edward Y. Villegas and Mario Velez},pdfsubject={Colciencias Young Researcher, Call 566},pdfkeywords={Hardness, Ab initio methods, Material design, Probabilistic ionic substitution, Evolutionary algorithms.},pdfpagemode=UseOutlines,bookmarks,bookmarksopen,pdfstartview=FitH, colorlinks,linkcolor=blue,urlcolor=blue,citecolor=blue]{hyperref}
\usepackage{authblk}
\usepackage{amsmath, amsthm, amssymb}
\usepackage{graphicx}
\usepackage{caption}
\usepackage{subcaption}
\usepackage{url}
\usepackage[numbers,sort&compress]{natbib}

\renewcommand{\citet}[1]{\citeauthor{#1} \citeyearpar{#1}}
\renewcommand{\cite}[1]{\citep{#1}}

\author{Edward Y. Villegas\thanks{email: \href{mailto:evillega@eafit.edu.co}{evillega@eafit.edu.co}}}
\author{Mario V\'elez}
\affil{Departamento de Ciencias B\'asicas, Universidad EAFIT, Medell\'in - Colombia}
\title{A new theoretical superstoichiometric hard $TiN_x$ crystal}
\date{}

\begin{document}

\maketitle

\addcontentsline{toc}{section}{Abstract}
\begin{abstract}
In this article we present the application of a \textit{ab initio} methodology to design new hard materials with crystal behavior. In special, we applied this methodology in the search of new materials of the \(TiN_x\) family. This search will find a set of structures super-stoichiometrics \(Ti_3N_4\) and properties of the candidate to greatest hardness are evaluated. Hardness model used in this work is described by \citet{Simunek2006} \cite{Simunek2006}.
\end{abstract}


\section{Introduction}
Today, material and energy industry search innovative strategies based on computational methods and high performance computing, whose can explore nature in atomic lenght scale \cite{Hergert2004,Karakasidis2007}. Discover of new inorganics compounds is a critical factor in technology developments, however this searching process is commonly very slow and highly empirical \cite{Hautier2011}.

The problem of crystal composition and structure prediction can be approximate as an optimization problem. An optimization algorithm is used to find structural parameters and atomic sites minimizing the total energy obtained by \textit{ab initio} methods. This approximation requires high computational cost \cite{Glass2006,Hautier2011,Oganov2009}.

This optimization problem exhibe great difficulties due the high dimensionality of coordinate space (\(3N+3\), where \(N\) is number of atoms in the system), high roughness in the response surface of free energy, energy sensibility to minor changes in interatomic distances, redundancy in the search spaces (because crystals are formed by unitary cells) and additional high computational cost of \textit{ab initio} calculations \cite{Glass2006, Hautier2011}.

Due to the characteristics of the problem, is suggested the use of evolutionary algorithms to sampling the search spaces \cite{Oganov2006,Glass2006}. This model is explained in \citet{Oganov2006}\cite{Oganov2006} and implemented in the code \textit{USPEX} \cite{Glass2006}, in which ramdom sample of crystals are building using symmetry criteria and are evaluated to find best total energies (or others properties), and then over the best candidates are applied evolutionary operators to enhance properties in next generations of crystals. However, not any random parameter with initial conditios of configuration file produce a first generation acceptable.

An alternative solution to this problem is using a data mining methodology which has less computational cost to produce new candidates to possible stable materials. Using an statistical model with a big database of inorganic compounds, chemical rules are extracted and are used with specific atomic species \cite{Hautier2011}.

\citet{Hautier2011} developed a probabilistic model to quantify and predict new posibble crystals based on the empirical model of ionic substitution of Goldschmidt (1926) \cite{Hautier2011}. This model obtain the probability of a given substitution could ocurre in the nature, using as the sample a big database of crystal structures. When this probability is greater than a certain threshold, we can accept the substitution as possible and obtain a new candidate material. Probabilistic ionic substitution is implemented in the online API and cloud service \textit{Materials Project}\cite{Hautier2011} in \textit{Structure Predictor}\cite{Jain2013} module.

We combined both alternatives in one progressive methodology, in which data mining approach is used as a guest for composition and structural parameters in evolutionary algorithm, thus probability of failure in first generation of structure reduce. This article applied the methodology to the case of study of \(TiN_x\) family. The goal of this progressive methodology is not only predict new materials (data mining approach) also obtain properties materials optimization with low computational cost.

\section{Computational detail}

Probabilistic ionic substitution is applied through the cloud service of \textit{Materials Project} with the \textit{Structure Predictor} module to nitrogen and titanium atoms. These selection of atoms correspond to an interest in possible new hard materials in the family of \(TiN_x\) compounds.

Results of running this data mining procedure, are showed in table \ref{tbl:strpre}.

\begin{table}
  \caption{Structure Predictor running with nitrogen and titanium atoms.}
  \label{tbl:strpre}
  \begin{tabular}{lll}
    \hline
    Oxidation state & Number of candidates & Computing time (hh:mm) \\
    \hline
    N:[-3], Ti:[2] & 0 & 0:51\\
    N:[-3], Ti:[3] & 6 & 0:43\\
    N:[-3], Ti:[4] & 9 & 1:00\\
    N:[-3], Ti:[2, 3] & 0 & 5:48\\
    N:[-3], Ti:[3, 4] & 0 & 3:06\\
    N:[-3, 1], Ti:[2] & 0 & 3:05\\
    N:[-3, 1], Ti:[3] & 0 & 14:18\\
    N:[-3, 1], Ti:[2, 3, 4] & 0 & 1:14\\
    N:[-3, 3, 5], Ti:[3, 4] & 0 & 1:00\\
    \hline
  \end{tabular}

\end{table}

\begin{figure}
 \centering
 \begin{subfigure}[b]{0.4\textwidth}
  \includegraphics[width=\textwidth]{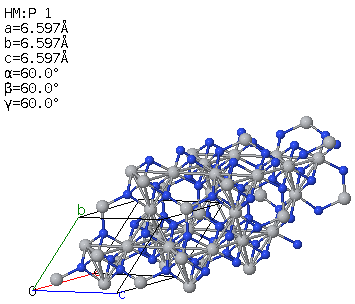}
  \caption{\textit{Fd\={3}m} spinel.}
  \label{img:spinel_mp}
 \end{subfigure}
 \quad
 \begin{subfigure}[b]{0.4\textwidth}
  \includegraphics[width=\textwidth]{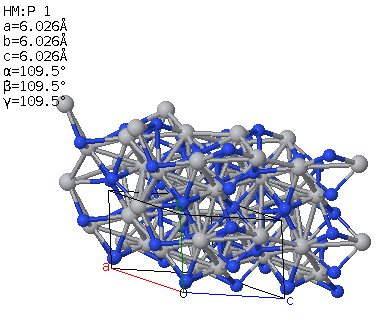}
  \caption{\textit{I\={4}3d}.}
  \label{img:i_mp}
 \end{subfigure}

 \begin{subfigure}[b]{0.4\textwidth}
  \includegraphics[width=\textwidth]{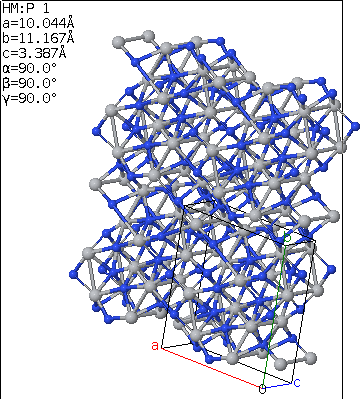}
  \caption{\textit{P1}.}
  \label{img:p1_mp}
 \end{subfigure}
 \quad
 \begin{subfigure}[b]{0.4\textwidth}
  \includegraphics[width=\textwidth]{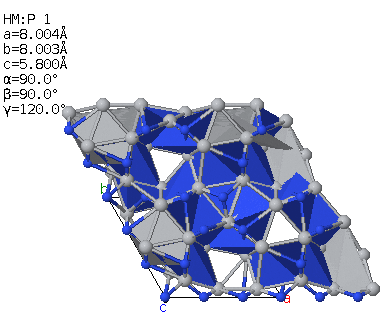}
  \caption{\textit{P31c}.}
  \label{img:p3_mp}
 \end{subfigure}

 \begin{subfigure}[b]{0.4\textwidth}
  \includegraphics[width=\textwidth]{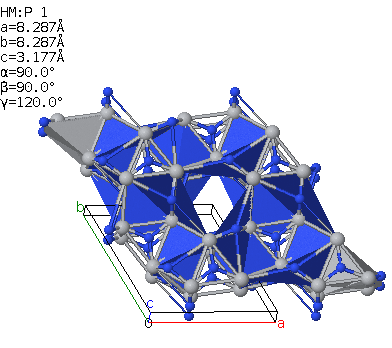}
  \caption{\textit{P63m}.}
  \label{img:p6_mp}
 \end{subfigure}

 \caption{\(Ti_3N_4\) by probabilistic ionic substitutions.} 
\end{figure}

We find two candidate compositions, 6 structures with relation 1:1 and 9 structures with composition \(Ti_3N_4\). The last composition contain teh following symmetry groups: \textit{Fd\={3}m} (Fig. \ref{img:spinel_mp}), \textit{I\={4}3d} (Fig. \ref{img:i_mp}), \textit{P1} (Fig. \ref{img:p1_mp}), \textit{P31c} (Fig. \ref{img:p3_mp}) y \textit{P63m} (Fig. \ref{img:p6_mp}).

Then, using these structure as guest in evolutionary algorithm, we find 94 stable structures in 9 generations. Here, we use Quantum Espresso\cite{Giannozzi2009} software to perform \textit{ab initio} calculations. Pseudopotentials ultrasoft building by Vanderbil method were used in \textit{ab initio} calculations, type Perdew-Zunger (LDA) with exchange-correlation energy approximation.

The electronic energy threshold criteria was $10^{-6}\quad Ry$ by unitary cell and the integration above Brillouin zone is performed in a Monkhorst-Pack grid of $36 \times 36 \times 54$. Gaussian occupation is setup to $0.136eV$ and system is in a low pressure condition. Band structure calculation is performed using 98 points over irreducible Brillouin contour.

Table \ref{tbl:uspex} illustrate distribution of crystal structures obtained using the evolutionary algorithm USPEX with optimization of hardness property. 

\begin{table}
 \centering
 \caption{Crystal phases generated with USPEX evolutionary algorithm.}
 \label{tbl:uspex}
 \begin{tabular}{|c|c|c|}
   \hline
   Crystal system & Symmetry group & Number of structures\\
   \hline
   Triclinic & $P1$ & 52\\
    & $P\overline{1}$ & 4\\
   \hline
    & $P2$ & 1\\
   Monoclinic & $P2_1$ & 2\\
    & $C2/c$ & 1\\
   \hline
   Orthorhombic & $Pmn2_1$ & 2\\
    & $Cmc2_1$ & 1\\
   \hline
    & $P4_22_12$ & 8\\
    & $P4nc$ & 8\\
   Tetragonal & $I\overline{4}c2$ & 3\\
    & $P4/nmm$ & 1\\
    & $P4_2/mcm$ & 2\\
    & $P4_2/nmc$ & 4\\
   \hline
    & $Im\overline{3}$ & 1\\
   Cubic & $Pm\overline{3}m$ & 3\\
    & $Im\overline{3}m$ & 1\\
   \hline
 \end{tabular}
\end{table}

\begin{figure}
 \centering
 \begin{subfigure}[b]{0.6\textwidth}
  \includegraphics[width=\textwidth]{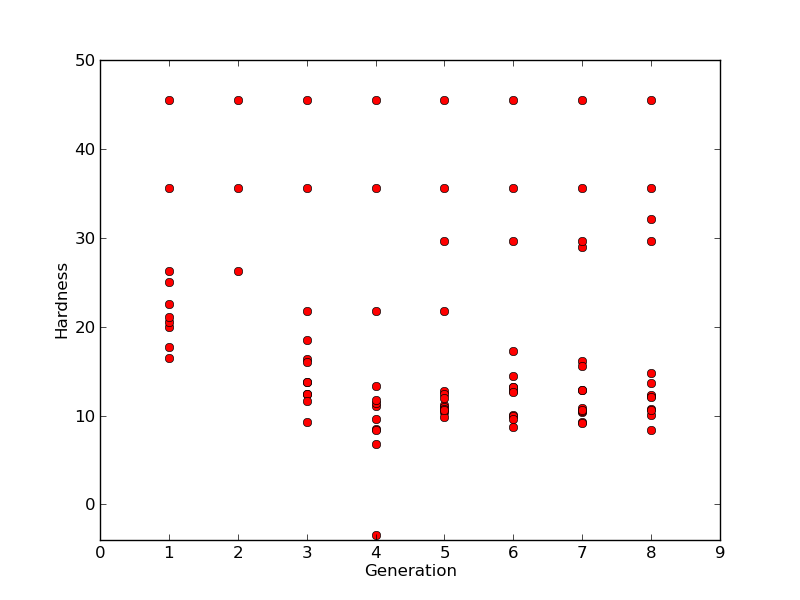}
  \caption{Hardness evolution.}
  \label{img:hard}
 \end{subfigure}
 \quad
 \begin{subfigure}[b]{0.6\textwidth}
  \includegraphics[width=\textwidth]{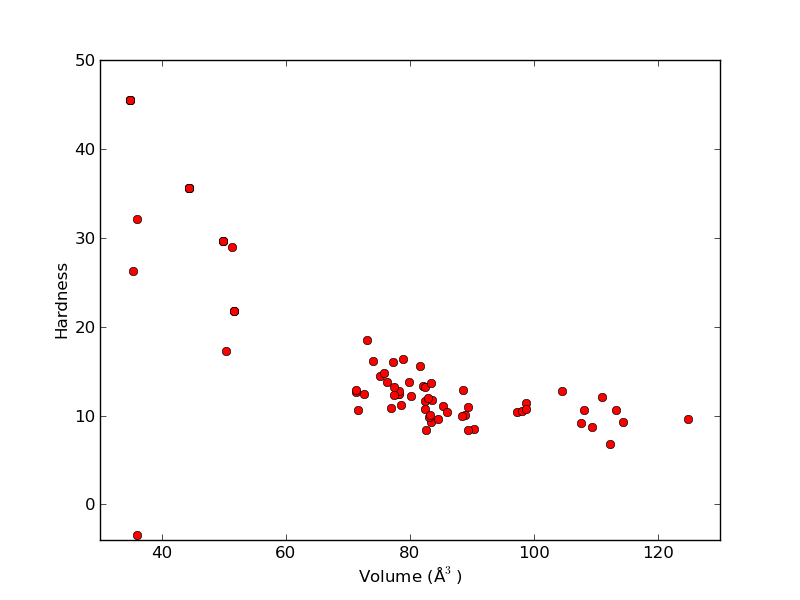}
  \caption{Relation between hardness and volume in structures generated by USPEX.}
  \label{img:hv}
 \end{subfigure}

 \caption{Hardness behavior in structures generated in USPEX.}
 \label{img:relations}
\end{figure}

Evolution of hardness property in generations and related with change in volume, is illustrated in figures \ref{img:relations}. The evolution of hardness in figure \ref{img:hard} show a strong candidate to high hardness since first generation, which is an structure with symmetry group \(P4_22_12\). Hardness relation with volume shows an inversed relation according to equation of \citeauthor{Gao2003} and \citet{Simunek2006}\cite{Gao2003,Gao2006,Simunek2006}. More compacts structures are more hardness.

This work use definition of intrinsic hardness by equations of \citet{Simunek2006},
\begin{eqnarray*}
e_{i} & = & \frac{Z_{i}}{R_{i}},\\ 
S_{ij} & = & \frac{\sqrt{e_{i}e_{j}}}{d_{ij}n_{ij}},\\ 
f_{e} & = & 1-\left[k\left(\prod_{i=1}^{k}e_{i}\right)^{1/k}/\sum_{i=1}^{k}e_{i}\right]^{2},\\
H & = & \frac{C}{V}n\left[\prod_{i,j=1}^{n}N_{ij}S_{ij}\right]^{1/n}\exp(-\sigma f_{e}), 
\end{eqnarray*}
where \(Z_i\) is the number of valence electrons of atom \(i\), \(R_i\) is the radii to integration of charge is neutral, \(n_{ij}\) is the number of bonds between atoms \(i\) and neighbour atoms \(j\) at distance \(d_{ij}\). Additional, \(N_{ij}\) is the number of binary systems with atoms \(i\) and \(j\) in the crystal, and \(k\) is the number of different atoms in the crystal. \(C\) and \(\sigma\) are experimental constants defined as typical values of 1550 and 4 for systems similars to diamond, silice (covalent behavior) and common salt (ionic bahavior).

\section{Results}

The phase \textit{Fd\={3}m} \(Ti_3N_4\) is also knowing as c-\(Ti_3N_4\) or spinel phase. This phase is obtained by substitution of any other mineral with spinel phase, e.g. magnetite \(Fe_3O_4\) and rescale lattice parameters (crystal coordinates are very similar). This spinel phase is a controversial theoretical phase predicted by \citet{Ching2000}\cite{Ching2000} and it is speculated as a transition phase in ammonolysis process of solution phase $Ti(NMe_2)_4$\cite{Dubois1994,Jackson2006}. \textit{Ab initio} calculations by \citet{Ching2000,Kroke2004} indicate a possible semiconductor with bandgap of $0.25eV$ and high hardness\cite{Ching2000,Kroke2004}. c-\(Ti_3N_4\) is the unique phase mentioned in literature with composition \(Ti_3N_4\)\cite{Dubois1994,Ching2000,Kroke2004,Li2004,Jackson2006}.

We find using data mining 8 more structures candidates with the same stoichiometry, used as guest in evolutionary algorithm, in which 94 crystal structures are generated as stable structures.

\begin{figure}
 \centering
 \includegraphics[scale=0.5]{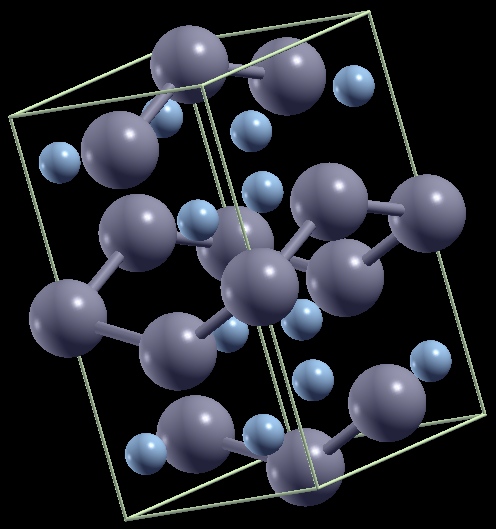}
 \caption{Optimized structure to superhardness.}
 \label{img:opt}
\end{figure}

The new theoretical hardness super-stoichiometric titanium nitride of phase \(P4_22_12\) has total energy of -856.863483 Ry and Fermi energy of 6.4204 eV, with total energy convergence of $2\times 10^{-8}\quad Ry$ by unitary cell. Lattice parameters and atom sites are showing in table \ref{tbl:opt} and plot associated in figure \ref{img:opt}.

\begin{table}
 \centering
 \caption{Optimized parameters of \(Ti_3N_4\) superhardness candidate. Position is in crystal coordinates.}
 \label{tbl:opt}
 \begin{tabular}{cc}
  \hline
  \multicolumn{2}{c}{Celda}\\
  \hline
  Crystal system & Cubic \\
  Symmetry group & $P4_22_12$ (94) \\
  a (\AA) & 2.79339 \\
  b (\AA) & 2.79339 \\
  c (\AA) & 4.46395 \\
  \hline
  \multicolumn{2}{c}{Asymmetrical unit} \\
  \hline
  N   &    0.000000   0.193090   0.851679\\
  Ti  &    0.000000   0.000000   0.500000\\
  Ti  &    1.000000   0.500000   0.120679\\
  \hline
 \end{tabular}
\end{table}

Following we show the strength tensor in units of \textit{kbar}
\begin{equation*}
 \begin{pmatrix}
  415.25 & 0 & 0\\
  0 & 415.24 & 0\\
  0 & 0 & 397.51
 \end{pmatrix}
\end{equation*}.

\begin{figure}
 \centering
 \includegraphics[scale=0.55]{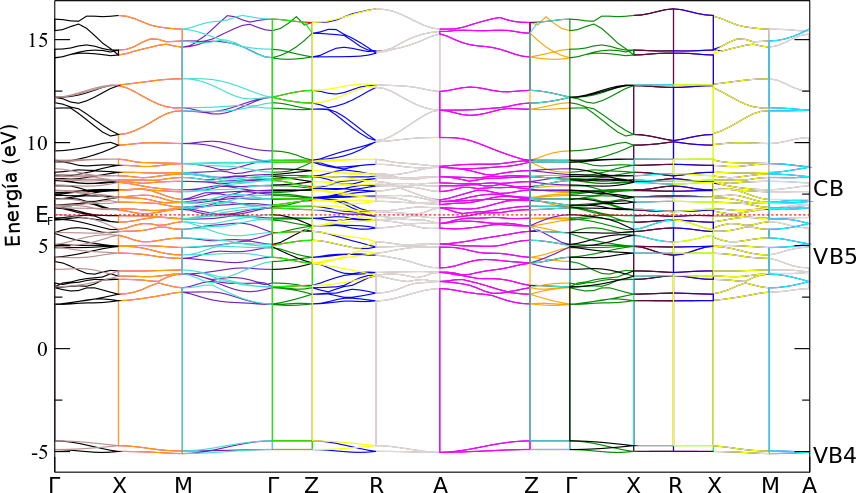}
 \caption{Band structure of \(Ti_3N_4\) candidate to superhardness.}
 \label{img:band_fermi}
\end{figure}

In figure \ref{img:band_fermi} band structure is illustrated, and we can see an overlapping around fermi energy of valence and conduction band, indicating a conductive behavior in this ceramic structure, a common characteristic with others titanium nitrides and carbon nitrides\cite{Ern1965,Zhang2012}.

In order to estimate hardness, we calculate also Lowdin population for the crystal, and the result is showing in sites of the asymmetrical unit in table \ref{tbl:lowdin}.

\begin{table}
 \centering
 \caption{Partial charge and nominal valence over asymmetric unit.}
 \label{tbl:lowdin}
 \begin{tabular}{ccc}
  \hline
  Atom & Partial charge & Nominal valence\\
  \hline
  Nitrogen & 0.70425 & -2\\
  Titanium 1 & -0.86625 & +2\\
  Titanium 2 & -0.9906 & +3\\
  \hline
 \end{tabular}
\end{table}

Using values obtained in this work, with formulae of \citeauthor{Simunek2006} and literature data to titanium nitride (1:1), we can compare an adimensional value of hardness associated with titanium nitride (1:1) equal to 5.7082 and a value of 83.9252 for titanium nitride \(Ti_3N_4\) \(P4_22_12\).

\section{Conclusions}

We find a new theoretical candidate super-stoichiometric titanium nitride \(P4_22_12\) to superhardness (and also electrical condutor) using a combined methodology of computational material design, searching guests by data mining approach and then optimized total energy and specific properties on the material with evolutionary algorithms.

Using the methodology we find more than one possible stable structures of \(Ti_3N_4\), in which we can find also the unique precedence to this work, the c-\(Ti_3N_4\). This observations agree with the work of \citeauthor{Ching2000}.

This methodology can be applied to diversity of properties, in a way to search specific material properties via \textit{ab initio} calculatios. Application of great interest in material industry.

\addcontentsline{toc}{section}{Acknowledgement}
\section*{Acknowledgement}
The authors thank to the Center of Scientific Computing APOLO of Univerisidad EAFIT and Juan David Pineda (systems engineer and system administrator of APOLO) for the contribution on computing resource.

Also, the authors thank (and posthumous memory) to Andriy Lyakhov, main developer of the USPEX code, who in cooperation we can port compatibility of USPEX code with octave instead of matlab.

\addcontentsline{toc}{section}{Supp. Info}
\section*{Supporting Information Available}

These results are part of the project ''Modelo computacional multiescala de nanoindentaci\'on para caracterizaci\'on de nanomateriales'' supported by the Administrative Department of Science, Technology, and Innovation of Colombia (COLCIENCIAS) and Universidad EAFIT on the 566 (J\'ovenes investigadores e innovadores 2012) call.

\addcontentsline{toc}{section}{References}
\bibliographystyle{unsrtnat}
\bibliography{evillega}

\end{document}